\documentstyle{article} 
\begin{document} 
\title{Non-regular Potentials and Sources for Static Axisymmetric
Electrovacs}  \author{L. Fern\'andez-Jambrina\\Departamento de F\'{\i}sica
Te\'orica II,\\Facultad de Ciencias F\'{\i}sicas,\\Universidad Complutense
\\E-28040-Madrid, Spain}\date{}   \maketitle
\begin{abstract}
In this lecture a new formalism for constructing electromagnetic surface sources
for static axisymmetric electrovacs is presented. The electrostatic and
magnetostatic sources are derived from the discontinuities of the scalar
potentials. This formalism allows the inclusion of two kinds of dipole sources:
Sheets of dipoles and the dipole moment of a distribution of monopoles. It is a
generalization of a previous formalism in order to cope with asymptotically
monopolar electric fields. \end{abstract}  

\section{Introduction}

The problem of finding compact physically reasonable material sources
that could be matched to an appropriate asymptotically flat vacuum
spacetime is a task of great relevance in general relativity.
Unfortunately, very few exact solutions are known for the stationary axisymmetric
case (cfr. \cite{Esc} for a recent review). 

 Instead of considering volume sources this lecture will be devoted to the
calculation of electromagnetic sources for static axisymmetric asymptotically
flat electrovacs. The electric charge distribution is easily obtained from the
integration of the Maxwell equations \cite{is1}, but electric and magnetic
moment distributions cannot be calculated in that way. In order to achieve that
goal a generalization of the approach followed in \cite{second}, \cite{tesis}
will be attempted. In these references magnetic and electric sources for
static electrovacs were constructed thanks to the introduction of an
asymptotically cartesian coordinate $Z$, provided that the electric field was
not monopolar. A way to circumvent that difficulty will be shown here. This will
also allow to calculate the contribution of the charge density to the electric
dipole. 

In the next section the Green identity will be used to rederive the classical
expression for the dipole surface density that arises from a discontinuous
scalar potential. This will be helpful to understand its generalization to
curved spacetimes in section \ref{shell}. An example of an application of this
formalism is presented in section \ref{example}. The results will be
discussed at the end.

\section{Non-relativistic dipole surface sources}

Let us consider a non-relativistic physical vector field, $E$, (electric or
magnetic) which can be obtained by differentiation of a scalar potential, $V$,
($E=-dV$) that fulfils the flat-spacetime Laplace equation. From the classical
theory of potential \cite{Kel} it is known that, if the field is discontinuous
across a surface $S$, (and therefore the normal derivative of $V$ is
discontinuous) then on crossing $S$ a layer of monopole charge is
encountered and the surface density thereof, $\sigma_1$, is given by:

\begin{equation}
\sigma_1=\frac{1}{4\pi}\,\left[E\cdot
n\right]=-\frac{1}{4\pi}\,\left[\frac{dV}{dn}\right] 
\end{equation}
where $n$ is the outer unitary normal to $S$ and a square bracket denotes
the difference ($[a]=a^+-a^-$) between the values of a quantity on
the outer ($a^+$) and inner ($a^-$) sides of $S$.

If, besides, not only the field but also the potential is discontinuous on $S$,
then the dipole density on $S$ can be constructed in the following way: Since
both the cartesian coordinate $z$ and the scalar potential $V$ satisfy the
Laplace equation out of $S$, then the Green identity is valid on ${\bf
R}^3_+\cup{\bf R}^3_-$, that is, the euclidean space outside and inside the
surface $S$:

\begin{eqnarray}
 0&=&\int_{{\bf
R}^3_+\cup{\bf R}^3_-}\,d^3x\,(V\,\Delta z-z\,\Delta
V)=\nonumber\\&=&\int_{S^2(\infty)\cup S^-}\,dS\,(
V\,\frac{dz}{dn}-z\,\frac{dV}{dn})-\int_{ S^+}\,dS\,(
V\,\frac{dz}{dn}-z\,\frac{dV}{dn})\label{Green} \end{eqnarray}
since the boundary of ${\bf R}^3_+$ consists of the sphere at infinity and $S$
and the boundary of ${\bf R}^3_-$ is just $S$.

Taking into account that the asymptotic behaviour of $V$ is known from its
expansion in Legendre polynomials and inverse powers of the spherical radius,
the integral at infinity can be performed and the other terms can be identified
as the dipole surface density $\sigma_2$ on $S$:

\begin{equation}
\sigma_2=\frac{1}{4\,\pi}\,\left\{n\cdot
u_z\,[V]-z\,\left[\frac{dV}{dn}\right]\right\}\label{clasden}
\end{equation}

The first term in this expression arises as the contribution of a sheet
of dipoles \cite{Kel} whilst the second one is the moment density of the
$\sigma_1$ distribution. 

\section{Relativistic thin shells}\label{shell}

In this section a way of generalizing the expression
for the dipole density to Maxwell fields in curved spacetimes will be
introduced. The metric for the static axially symmetric
electromagnetic-gravitational system can be written in Weyl coordinates:

\begin{equation}ds^{2}=-e^{2U}\,dt^{2}+e^{-2U}\{e^{2k}(d\rho^{2}+dz^{2})+\rho^2
d\phi^{2}\}\label{eq:can}
\end{equation}
where $U$, $k$ are functions of $\rho$ and $z$.

The scalar potential $V$ for either the electric or the magnetic field satisfies
the following equation, that can be derived from Maxwell's vacuum equations only,
even if the electromagnetic stress tensor is not the source of the gravitational
field:

\begin{equation}
\frac{1}{\sqrt{g}}\,\partial_\mu\,\left\{\sqrt{g}\,e^{-U}\,g^{\mu\nu}\,
\partial_\nu\, V\right\}=0\label{V} \end{equation}
where the metric $g$ is the one induced by (\ref{eq:can}) on each of the
hypersurfaces $t=const.$

Hence the results will be valid also for Maxwell fields in the geometry defined
by (\ref{eq:can}).

Our aim will be to cope with compact sources, therefore only asymptotically
flat metrics will be considered: 

\begin{eqnarray}ds^2\simeq-(1-\frac{2m}{r})\,dt^2+
(1+\frac{2m}{r})\{dr^2+(r^2+\alpha\,r)(d\theta^2 +\sin^2\theta
d\phi^2)\} \end{eqnarray} 
in terms of the total mass $m$ and a constant $\alpha$ in a coordinate patch
described by $\{t,\phi,r,\theta\}$. On the other hand, the scalar potential
will be required to have the following expansion, where $Q$ is the total
monopole charge, $d$ is the total dipole moment and $\beta$ is another constant:

\begin{equation}
V=\frac{Q}{r}+\frac{d\,\cos\theta}{r^2}+\frac{\beta}{r^2}+O(r^{-3})
\end{equation}

If the source for the field is located on a surface $S$, then equation
(\ref{V}) can be integrated on the regions $V_3^+$ and $V_3^-$ in which the
hypersurfaces $t=const.$ are split by $S$. Since the integrand is a total
derivative, the integral can be reduced to a surface integral on $S$ and
another on the sphere at infinity and therefore it yields the expression for the
monopole charge density:

\begin{equation}
\sigma_1=-\frac{1}{4\pi}\,e^{-U}\,\left[\frac{dV}{dn}\right]\label{charden}
\end{equation}
which is the formula that was obtained by Israel in \cite{is1}, but written in
terms of the potential instead of the field. The only difference with the
non-relativistic situation is the appearance of a metric factor.

After the fashion of \cite{second}, \cite{first}, \cite{zip}, an asymptotically cartesian
function $Z$ will be introduced and will be taken to satisfy the same
differential equation as $V$:

\begin{equation}
\frac{1}{\sqrt{g}}\,\partial_\mu\,\left\{\sqrt{g}\,e^{-U}\,g^{\mu\nu}\,
\partial_\nu\, Z\right\}=0\label{Z} \end{equation}

Hence a Green identity can be used to reduce the following integral on the two
regions $V_3^+$ and $V_3^-$:

\begin{eqnarray}
0&=&\int_{V_3^+\cup
V_3^-}\sqrt{g}\,\frac{1}{\sqrt{g}}\,\left\{Z\,\partial_\mu\,(\sqrt{g}\,e^{-U}
\,g^{\mu\nu}\, \partial_\nu\,
V)-\right.\nonumber\\&-&\left.V\partial_\mu\,(\sqrt{g}\,e^{-U}\,g^{\mu\nu}\,
\partial_\nu\, Z)\right\}\,dx^1dx^2dx^3= \nonumber\\&=&\int_{\partial
V_3^+\cup\partial
V_3^-}\,dS\,e^{-U}\left(Z\,\frac{dV}{dn}-V\,\frac{dZ}{dn}\right)
\end{eqnarray} 

Assuming that $V$ may be discontinuous on $S$ and calculating the integral at
infinity with the information given by the asymptotic expansions, the
following expression is obtained:

\begin{equation}
0=-4\,\pi\,d+\int_S\,dS\,e^{-U}\,\left[V\,\frac{dZ}{dn}-Z\,
\frac{dV}{dn}\right]
\end{equation}

This equation yields the expression for the dipole density on $S$, as it
happened in the non-relativistic case, in terms of the discontinuities of $V$:

\begin{equation}
\sigma_2=\frac{1}{4\pi}\,e^{-U}\,\left[V\,
\frac{dZ}{dn}-Z\,\frac{dV}{dn}\right]
\label{magden}
\end{equation}

This formula is again similar to the classical one (\ref{clasden}),
except for a metric factor. As it happened then, there is a contribution of a
sheet of dipoles and also of the distribution of monopoles. This last term was
not considered in \cite{second}.

\section{An example: Bonnor's magnetic dipole}\label{example}

 As an example of how this formalism works, the magnetic source for Bonnor's
magnetic dipole \cite{Bon2} will be calculated. This solution is the Bonnor
transform of the Kerr metric \cite{Bon1}:

\begin{eqnarray}
ds^2=-(1-\frac{2mr}{r^2-a^2\cos^2\theta})^2dt^2+\nonumber\\+
(1-\frac{2mr}{r^2-a^2\cos^2\theta})^{-2}\left\{
 (r^2-a^2-2mr)\sin^2\theta d\phi^2+\right.\nonumber\\
\left.
\frac{(r^2-a^2\cos^2\theta-2mr)^4}{[(r-m)^2-(a^2+m^2)\cos^2\theta]^3}
(d\theta^2+\frac{dr^2}{r^2-2mr-a^2})\right\}
\end{eqnarray}

\begin{equation}
V=\frac{2am\cos\theta}{r^2-a^2\cos^2\theta}
\end{equation}
which describes the field around a magnetic dipole of mass equal to $2m$ and
magnetic moment $2am$.

As it was done in \cite{second} the radial coordinate will be taken to be
nonnegative. Events on the surface $r=0$ with collatitude $\theta$ will be
identified with those with $\pi-\theta$ and therefore the range of $\theta$
will be restricted to $[0,\pi/2)$ to avoid double-counting them. This means
that the magnetic potential is discontinuous on $r=0$:

\begin{equation}
[V]=-\frac{4m}{a\,\cos\theta}
\end{equation}

Since the solution satisfies the required asymptotic behaviours, only a $Z$
function satisfying (\ref{Z}) is needed for constructing the magnetic source:

\begin{equation}
Z=(r-3\,m)\,\cos\theta-\frac{2\,a^2\,m\,\cos^3\theta}
{r^2-a^2\cos^2\theta}
\end{equation}

Applying (\ref{charden}) and (\ref{magden}) to this solution, the expressions
for the magnetic density are obtained:

\begin{equation}
\sigma_2=\frac{m}{\pi\,a^4}\ \frac{|a^2\,\cos^2\theta-m^2\,\sin^2\theta|^{3/2}
}{\cos^4\theta}
\end{equation}

Obviously there is no monopole density. Integrating $\sigma_2$ on the surface
$r=0$, the correct result for the magnetic dipole moment is obtained:

\begin{equation} d=\int_S\ \sigma_2\ dS=\int^{2\pi}_{0}\int^{\pi/2}_{0}d\theta\
(m\,a\sin\theta)=2\,m\,a \end{equation}

This is the same result that was obtained in \cite{second} since there is no
contribution from monopole sheets, as was to be expected.

\section{Discussion}

It has been shown in this lecture a new method for constructing electromagnetic
sources for static axially simmetric spacetimes. With this new approach the
inclusion of asymptotically monopolar electric fields has been achieved. This
was not possible in a previous formalism \cite{second} because vector potentials
were used and therefore Dirac string singularities would be present if 
monopoles  were included. Also the influence of the charge distribution on the
dipole density has been considered. A further generalization to stationary
nonstatic axisymmetric spacetimes \cite{prep} is in preparation.

\noindent {\it The present work has been supported in
part by DGICYT Project PB92-0183; L.F.J. is supported by a FPI Predoctoral
Scholarship from Ministerio de Educaci\'{o}n y Ciencia (Spain). The author
wishes to thank F. J. Chinea, L. M. Gonz\'alez-Romero and J. A. Ruiz Mart\'{\i}n
for valuable discussions.}


\begin{thebibliography}{99}  
\bibitem{Esc} {\it El Escorial Summer School on Gravitation and General
Relativity 1992: Rotating Objects and Other Topics\/} (eds.: F. J. Chinea and
L. M. Gonz\'alez-Romero), Springer-Verlag, Berlin-New York (1993)
\bibitem{is1} W. Israel, {\it Phys. Rev.} {\bf D2}, 641 (1970) 
\bibitem{second} L. Fern\'andez-Jambrina and F.J. Chinea, {\it Class. Quantum
Grav.} {\bf 11}, 1489 (1994) [arXiv: gr-qc/0403118]
\bibitem{tesis} L. Fern\'andez-Jambrina {\it Ph.D. thesis}, Universidad Complutense de Madrid
(1994) ISBN: 84-699-0403-4 
\bibitem{Kel} O. D. Kellogg: {\em Foundations of Potential Theory.\/} Dover, New
York, 1954
\bibitem{first} L. Fern\'andez-Jambrina, F. J. Chinea, {\it Phys. Rev.
Lett.} {\bf 71}, 2521 (1993) [arXiv: gr-qc/0403102]
\bibitem{zip} L. Fern\'andez-Jambrina, {\it Class. Quant. Grav.} {\bf 
11},  1483 (1994) [arXiv: gr-qc/0403113]
\bibitem{Bon2} W. B. Bonnor, {\it Z. Phys.} {\bf 190}, 444 (1966)
\bibitem{Bon1} W. B. Bonnor, {\it Z. Phys.} {\bf 161}, 439 (1961)
\bibitem{prep} L. Fern\'andez-Jambrina, [arXiv: gr-qc/0404008]
\end{thebibliography}
 \end{document}